\documentclass[conference]{IEEEtran}
\IEEEoverridecommandlockouts
\usepackage{cite}
\usepackage{amsmath,amssymb,amsfonts}
\usepackage{algorithmic}
\usepackage{graphicx}
\usepackage{textcomp}
\usepackage{xcolor}
\usepackage{todonotes}
\def\BibTeX{{\rm B\kern-.05em{\sc i\kern-.025em b}\kern-.08em
    T\kern-.1667em\lower.7ex\hbox{E}\kern-.125emX}}
\begin{document}

\newcommand\copyrighttext{%
  \footnotesize \textcopyright © 2019 IEEE. Personal use of this material is permitted. Permission from IEEE must be obtained for all other uses, in any current or future media, including reprinting/republishing this material for advertising or promotional purposes,creating new collective works, for resale or redistribution to servers or lists, or reuse of any copyrighted component of this work in other works.\\
  Conference: 2019 International Multi-Conference on Engineering, Computer and Information Sciences (SIBIRCON)}
\newcommand\copyrightnotice{%
\begin{tikzpicture}[remember picture,overlay]
\node[anchor=south,yshift=10pt] at (current page.south) {\fbox{\parbox{\dimexpr\textwidth-\fboxsep-\fboxrule\relax}{\copyrighttext}}};
\end{tikzpicture}%
}

\newpage

\title{Comparison of UNet, ENet, and BoxENet for Segmentation of Mast Cells in Scans of Histological Slices\\
\thanks{The work was performed as part of the state assignment of IIF UB RAS (subject No. AAAA-A18-118020590108-7) and supported by the RF Government Act No. 211, dated March 16, 2013 (agreement 02.A03.21.0006).}
}

\author{
\IEEEauthorblockN{Alexander Karimov}
\IEEEauthorblockA{\textit{Engineering School of ITTCS} \\
\textit{Ural Federal University}\\
Ekaterinburg, Russia \\
sanya1992AK@gmail.com}
\and
\IEEEauthorblockN{Artem Razumov}
\IEEEauthorblockA{\textit{Institute of Natural Sciences} \\
\textit{Ural Federal University}\\
Ekaterinburg, Russia \\
airplaneless@yandex.ru}
\and
\IEEEauthorblockN{Ruslana Manbatchurina}
\IEEEauthorblockA{\textit{Engineering School of ITTCS} \\
\textit{Ural Federal University}\\
Ekaterinburg, Russia \\
manbatchurina99@gmail.com}
\and
\IEEEauthorblockN{Ksenia Simonova}
\IEEEauthorblockA{\textit{Engineering School of ITTCS} \\
\textit{Ural Federal University}\\
Ekaterinburg, Russia \\
ksansi@yandex.ru}
\and
\IEEEauthorblockN{Irina Donets}
\IEEEauthorblockA{\textit{Engineering School of ITTCS} \\
\textit{Ural Federal University}\\
Ekaterinburg, Russia \\
ira-don1225@gmail.com}
\and
\IEEEauthorblockN{Anastasia Vlasova}
\IEEEauthorblockA{\textit{Laboratory of Translational Medicine and Bioinformatics} \\
\textit{Institute of Immunology and Physiology}\\
Ekaterinburg, Russia \\
vlasova.9@mail.ru}
\and
\IEEEauthorblockN{Yulia Khramtsova}
\IEEEauthorblockA{\textit{Laboratory of Immunophysiology and Immunopharmacology} \\
\textit{Institute of Immunology and Physiology}\\
Ekaterinburg, Russia \\
hramtsova15@mail.ru}
\and
\IEEEauthorblockN{Konstantin Ushenin}
\IEEEauthorblockA{\textit{Institute of Natural Sciences} \\
\textit{Ural Federal University}\\
Ekaterinburg, Russia \\
konstantin.ushenin@urfu.ru}
}

\maketitle
\copyrightnotice

\begin{abstract}
Deep neural networks show high accuracy in the problem of semantic and instance segmentation of biomedical data. However, this approach is computationally expensive. The computational cost may be reduced with network simplification after training or choosing the proper architecture, which provides segmentation with less accuracy but does it much faster. In the present study, we analyzed the accuracy and performance of UNet and ENet architectures for the problem of semantic image segmentation. In addition, we investigated the ENet architecture by replacing of some convolution layers with box-convolution layers. The analysis performed on the original dataset consisted of histology slices with mast cells. These cells provide a region for segmentation with different types of borders, which vary from clearly visible to ragged. ENet was less accurate than UNet by only about 1-2\%, but ENet performance was 8-15 times faster than UNet one.
\end{abstract}


\section{Introduction}

Deep neural networks show high accuracy in the semantic and instance segmentation of biomedical data \cite{carneiro2017review}. However, deep neural networks are a computationally expensive approach in comparison with classical pattern recognition methods. Usually, high-quality automatic segmentation with deep neural networks requires a computational accelerator or a CPU with good computational performance. Thus, software engineers and biomedical engineers are forced to use cloud computing or add advanced computational power in the device for offline processing of data. This leads to an increase in the infrastructure annual cost or an increase in the main cost of developing biomedical devices.

Reducing the computational cost of deep learning may be achieved with two approaches. In the first approach, a deep neural network may be simplified after training. For example, this type of method is described in work \cite{han2015deep} and presented by pruning, trained quantization, and Huffman coding.

In another approach, several works propose neural network architectures that do not show the best accuracy of segmentation but consist of a significantly smaller number of trainable parameters than the more precise approaches. ENet \cite{paszke2016enet} is this type of architecture that has been proposed for real-time semantic segmentation.

In present work, we compare the accuracy and performance of three neural networks (UNet \cite{ronneberger2015u}, ENet \cite{ronneberger2015u}, and BoxENet) for semantic segmentation of histological slice scans. UNet was chosen for the current study as the gold standard for biomedical segmentation problems. ENet was chosen as the architecture that shows the moderate quality in segmentation but should show a high performance in image processing. In addition, we assume before the study that replacing the usual convolution with box convolution  (\cite{burkov2018deep}) may improve the quality of the segmentation. To test this idea, we implement BoxENet, which is ENet model from the original paper with a replacement of some bottleneck block to bottleneck blocks with box-convolutional layers \cite{burkov2018deep}.

For analysis, we chose an original dataset that included toluidine blue-stained histological slices with mast cells at different stages of degranulation. This dataset was manually segmented by an expert. The main reason for this choice is the ground truth segmentation with a region that had a different type of border: clear border, ragged border, and a cloud of points with no observable border. In particular, the last case is important because the neural network cannot learn on the cell membrane pattern \cite{dimopoulos2014accurate}. Let us briefly describe the dataset.

Mast cells are part of the human immune system. They are related to allergy reactions, and several deceases. Mast cells look substantially different at a different stage of their physiological role. Before the release of granules with histamine, these cells look like compact objects with a clearly visible border. Under some factor, the cells start to spread capsules with histamine. This process is named \textit{degranulation}. During degranulation, the cells' borders become unclear, and the cell surrounded by histamine capsules that a look like a cloud of points. At the end of the degranulation process, the original location of the cell is indistinguishable, and the cell is fully replaced by the cloud of points. At the final stage of the degranulation process, it is difficult to separate one cell from another.

\section{Methods}

\begin{table}[t]
\begin{tabular}{cc}
\hline
\multicolumn{1}{|c|}{\textbf{ENet}}                     & \multicolumn{1}{c|}{\textbf{BoxENet}}             \\ \hline
Downsampler(3$\rightarrow$16)                         & Downsampler(3$\rightarrow$16)                   \\
Bottleneck(16$\rightarrow$64, downsample)             & Bottleneck(16$\rightarrow$64, downsample)       \\
Bottleneck(64$\rightarrow$64)                         & Bottleneck(64$\rightarrow$64)                   \\
\textbf{Bottleneck(64$\rightarrow$64)}                & \textbf{BottleneckBoxConv(64$\rightarrow$64)}   \\
Bottleneck(64$\rightarrow$64)                         & Bottleneck(64$\rightarrow$64)                   \\
\textbf{Bottleneck(64$\rightarrow$64)}                & \textbf{BottleneckBoxConv(64$\rightarrow$64)}   \\
Bottleneck(64$\rightarrow$128, downsample)            & Bottleneck(64$\rightarrow$128, downsample)      \\
Bottleneck(128$\rightarrow$128)                       & Bottleneck(128$\rightarrow$128)                 \\
\textbf{Bottleneck(128$\rightarrow$128, dilation=2)}  & \textbf{BottleneckBoxConv(128$\rightarrow$128)} \\
Bottleneck(128$\rightarrow$128)                       & Bottleneck(128$\rightarrow$128)                 \\
\textbf{Bottleneck(128$\rightarrow$128, dilation=4)}  & \textbf{BottleneckBoxConv(128$\rightarrow$128)} \\
Bottleneck(128$\rightarrow$128)                       & Bottleneck(128$\rightarrow$128)                 \\
\textbf{Bottleneck(128$\rightarrow$128, dilation=8)}  & \textbf{BottleneckBoxConv(128$\rightarrow$128)} \\
Bottleneck(128$\rightarrow$128)                       & Bottleneck(128$\rightarrow$128)                 \\
\textbf{Bottleneck(128$\rightarrow$128, dilation=16)} & \textbf{BottleneckBoxConv(128$\rightarrow$128)} \\
Bottleneck(128$\rightarrow$128)                       & Bottleneck(128$\rightarrow$128)                 \\
\textbf{Bottleneck(128$\rightarrow$128, dilation=2)}  & \textbf{BottleneckBoxConv(128$\rightarrow$128)} \\
Bottleneck(128$\rightarrow$128)                       & Bottleneck(128$\rightarrow$128)                 \\
\textbf{Bottleneck(128$\rightarrow$128, dilation=4)}  & \textbf{BottleneckBoxConv(128$\rightarrow$128)} \\
Bottleneck(128$\rightarrow$128)                       & Bottleneck(128$\rightarrow$128)                 \\
\textbf{Bottleneck(128$\rightarrow$128, dilation=8)}  & \textbf{BottleneckBoxConv(128$\rightarrow$128)} \\
Bottleneck(128$\rightarrow$128)                       & Bottleneck(128$\rightarrow$128)                 \\
\textbf{Bottleneck(128$\rightarrow$128, dilation=16)} & \textbf{BottleneckBoxConv(128$\rightarrow$128)} \\
Upsampler(128$\rightarrow$64)                         & Upsampler(128$\rightarrow$64)                   \\
Bottleneck(64$\rightarrow$64)                         & Bottleneck(64$\rightarrow$64)                   \\
Bottleneck(64$\rightarrow$64)                         & Bottleneck(64$\rightarrow$64)                   \\
Upsampler(64$\rightarrow$16)                          & Upsampler(64$\rightarrow$16)                    \\
Bottleneck(16$\rightarrow$16)                         & Bottleneck(16$\rightarrow$16)                   \\
ConvTranspose2d(16$\rightarrow$2)                     & ConvTranspose2d(16$\rightarrow$2)              
\end{tabular}\caption{The implemented ENet and BoxENet architectures. The second architecture repeats the first, but bottleneck blocks with the dilated convolutions are replaced to bottleneck blocks with box-convolution layers.}\label{tbl:architectures}
\end{table}

\begin{figure}[!h] 
\includegraphics[width=0.45\textwidth]{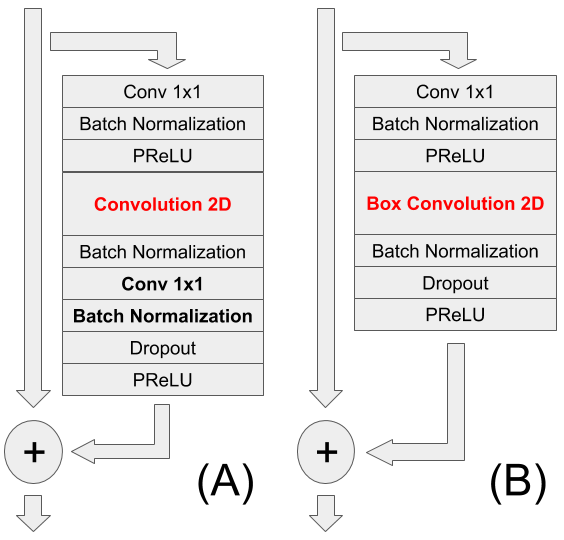}
\caption{Bottleneck blocks that are used for the current work in ENet (A) and BoxENet (B) neural networks. Red and bold fonts indicate the differences. }\label{fig:bottlenecks}
\end{figure}

\subsection{Dataset}

The mast cell dataset was acquired from histology preparations from previous studies, and laboratory animals were not purposefully processed for this study. The conditions of the housing and treatment of the animals during the experiment complied with the Directive of the European Parliament and the Council of September 22, 2010, on the protection of animals used for scientific purposes (2010/63/EU). Moreover, the study was approved by the local ethical committee of the Institute of Immunology and Physiology of the Ural Branch of the Russian Academy of Sciences (IIF UB RAS) (protocol No. 10, dated April 3, 2017).

The slices were obtained during a histological examination of the seminal vesicles and appendages of male Wistar rats of different ages. The preparations were fixed in 10\% formalin solution for 1 day. After standard tissue processing on a Shandon Excelsior closed-type automatic machine (MICROM International GmbH, Germany), the material was embedded in paraffin using an EG 1160 paraffin embedding station for biological tissue (Leica, Germany). Next, the paraffin blocks were cut on a semi-automatic microtome Thermo Microm HM 450 (MICROM International GmbH, Germany). The slice thickness was 4 $\mu$m. Then, the preparations were scanned with a light optical microscope (Leica, Germany).

The full dataset includes 168 images with a resolution of 1024 x 1280 px and standard RGB channels. Each image consists of 5 to 7 mast cells. About 40\% of all cells appear close to each other and have joined borders. Images were cut on 244 tiles with 256 x 256 tiles accordingly to input the size of the neural networks. Ground truth segmentation of the dataset was performed manually by an expert. Separation of the full dataset on test and validation part was performed in 74:51 ratio and a random shuffling of the samples.

\subsection{Artificial Neural Networks}

For the present purposes, we compared three architectures: UNet \cite{ronneberger2015u}, ENet \cite{ronneberger2015u}, and an ENet modification (BoxENet), where bottleneck block with the standard convolutional layer replaced to the bottleneck block with box-convolution layers \cite{burkov2018deep}.

UNet \cite{ronneberger2015u} has become a gold standard for comparison of biomedical segmentation accuracy and performance. UNet \cite{ronneberger2015u} is a convolutional autoencoder with additional connections between the encoder and the decoder parts. This type of neural network also is named "an hourglass architecture". UNet and neural networks that are used the same approach show highly accurate results in a wide area of biomedical applications. In addition, this family of architectures works faster than approaches that are based on the pixel-wise classification \cite{ronneberger2015u}.

ENet architecture \cite{paszke2016enet} was developed for the real-time semantic segmentation problem. In total, the ENet from the original paper contains twice less trainable parameters than UNet. This neural network uses a convolutional block with short skipped connections, which is also named as a bottleneck block (see. Fig. \ref{fig:bottlenecks}). Also, this architecture used dilated convolutions, asymmetric convolutions, and other approaches that affect the perceptive field and the distribution of weights in the deep layers.  

Besides that, we implement ENet architecture with the replacement of some bottleneck blocks with usual convolution layers to bottlenecks block with box-convolution layers proposed in \cite{burkov2018deep}. From our assumption, this replacement may improve the quality of the segmentation. Variation of ENet architectures and replaced layers for BoxENet are presented in Fig. \ref{tbl:architectures}. Fig. \ref{fig:bottlenecks} shows the differences between the usual bottleneck block and the bottleneck block with box convolution layer.

All neural networks from the original paper were implemented with minor changes. The training was performed using dice similarity coefficient (DSC)  loss and ADAM optimizer with a learning rate of 0.0001. We did not apply the overlapping strategy and weight map that is usually used for the Unet architecture \cite{ronneberger2015u}. This unification of training approaches was performed to focus on the primary goal of the broader study because a more wide study is required to analyze the effect of other losses and weight maps on the accuracy of each neural network.

\section{Results}

\begin{figure}[!h] 
\includegraphics[width=0.45\textwidth]{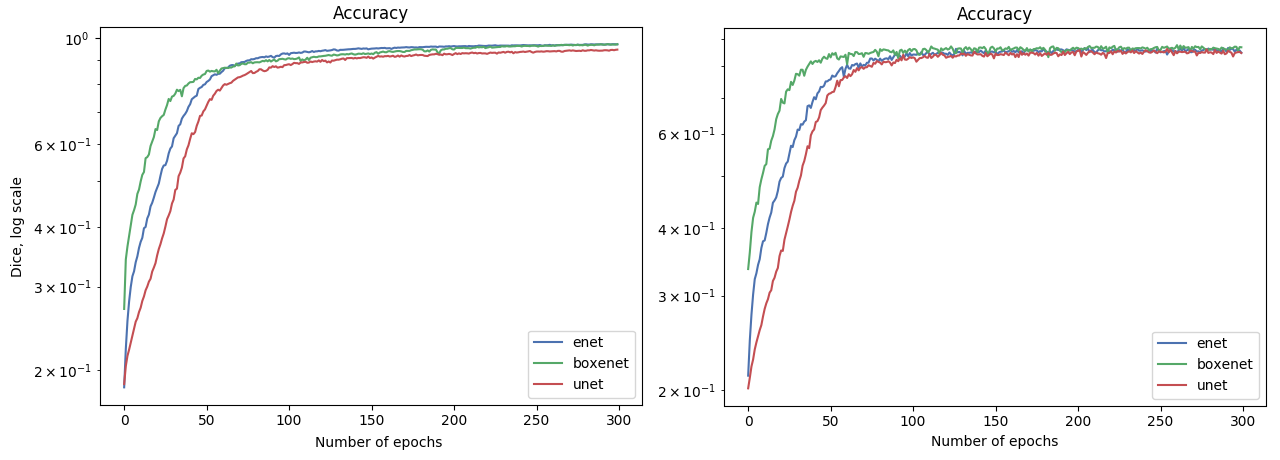}
\caption{Accuracy by epochs of the neural network training in terms of the DSC. The left image is for the training dataset, and the right image is for the validation dataset.}\label{fig:accuracy}
\end{figure}

As shown in Fig. \ref{fig:accuracy}, the ENet, and BoxENet accuracy increase faster than the UNet accuracy during the neural network training process. The best neural network accuracy is reached before 300 epochs for each of the three architectures.

\begin{figure}[!h] 
\includegraphics[width=0.45\textwidth]{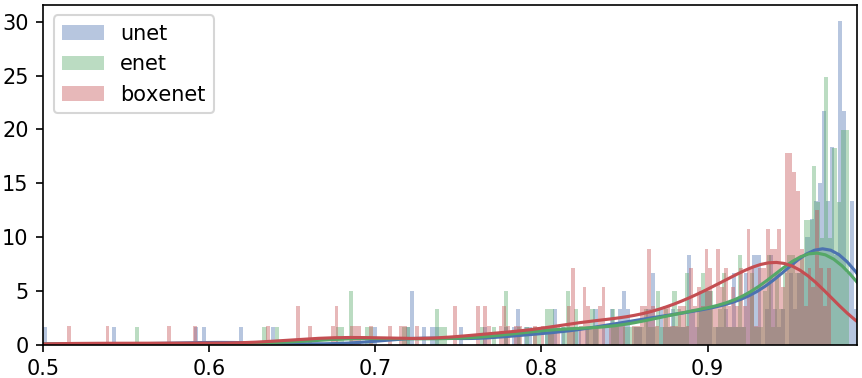}
\caption{Distribution of the DSC over images from the validation dataset.}\label{fig:hist}
\end{figure}

\begin{table}[h]
\begin{tabular}{l|lll}
              & UNet   & ENet   & BoxENet  \\ \hline
DSC coef. (mean)     & \textbf{0.9159} & 0.9107 & 0.8859 \\
DSC coef. (median)   & \textbf{0.9514} & 0.9469 & 0.9157 \\
DSC coef. (max)      & \textbf{0.9909} & 0.9851 & 0.9741 \\
DSC coef. (percentile 10) & \textbf{0.8075} & 0.7966 & 0.7671 \\ \hline
IoU (mean)     & \textbf{0.8538} & 0.8445 & 0.8092 \\
IoU (median)   & \textbf{0.9079} & 0.8996 & 0.8525 \\
IoU (max)      & 0.9757 & \textbf{0.9771} & 0.9571 \\
IoU (percentile 10) & \textbf{0.6744} & 0.6624 & 0.6161  \\ \hline
F1 (mean)     & \textbf{0.8509} & 0.8442 & 0.8076 \\
F1 (median)   & 0.8993 & \textbf{0.9023} & 0.8578 \\
F1 (max)      & \textbf{0.9785} & 0.9738 & 0.9487 \\
F1 (percentile 10) & \textbf{0.6713} & 0.6539 & 0.6207
\end{tabular}
\caption{Comparison of the neural network accuracy for the validation dataset. The bold text shows the highest value in each row.}\label{tbl:result}
\end{table}

Fig. \ref{fig:hist} and Table \ref{tbl:result} present a detailed comparison of the accuracy metrics on the validation dataset for the three neural networks. Accordingly to the table, UNet shows higher values in three metrics of the semantic segmentation accuracy. The median of the DSC distribution is lower for BoxENet in comparison with UNet and ENet as shown in Fig.\ref{fig:hist}. Therefore, the box convolutions did not sufficiently improve the accuracy of the ENet architecture.

\begin{figure}[!h] 
\includegraphics[width=0.45\textwidth]{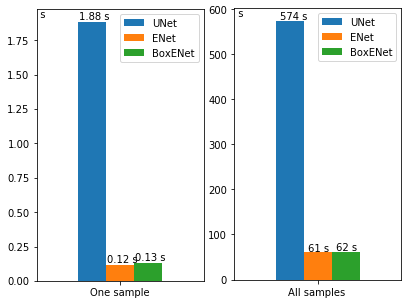}
\caption{Comparison of computational performance for three architectures on one sample and the full validation dataset.}\label{fig:performance}
\end{figure}

As shown in Fig. \ref{fig:performance}, the ENet, and BoxENet computational performance was up to 10 to 15 times higher than the UNet performance. That analysis were performed on a computational accelerator NVIDIA Tesla K80.

\begin{figure*}[!h] 
\includegraphics[width=\textwidth]{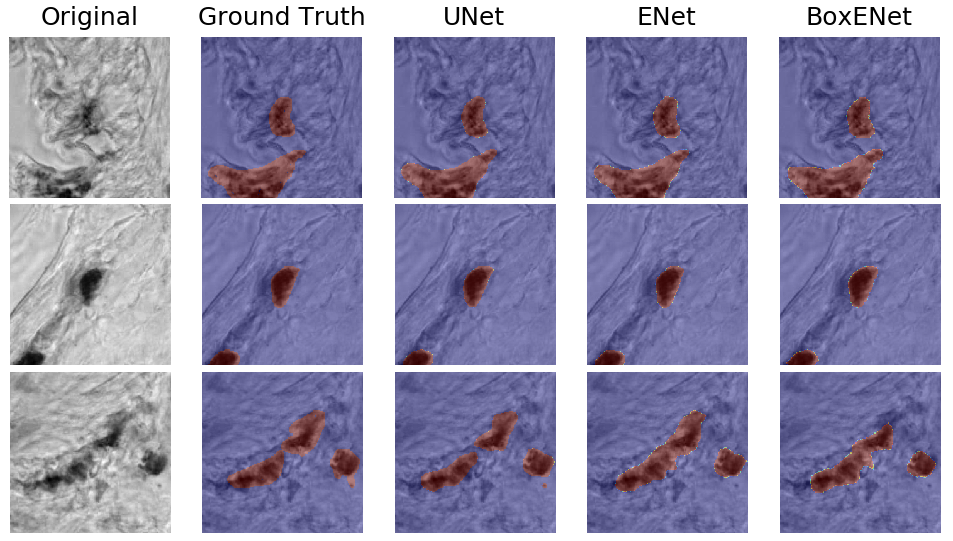}
\caption{Original image, ground truth, and semantic segmentation performed with three neural networks.}\label{fig:examples}
\end{figure*}

Figure \ref{fig:examples} shows segmentation results. Each of the three neural networks shows the correct segmentation for objects on the image border. All neural networks were insensitive to the stage of mast cell degranulation and correctly segmented cells with any border. However, UNet shows a better result in segregating closely located cells even without the application of the weighted map in training. That may lead to issues with the counting of objects on images, and complicate the instance segmentation with ENet instead of UNet.

\section{Discussion and Conclusion}

In the current study, we compared the accuracy and performance of UNet, ENet, and ENet with box convolutional layers (BoxENet). The comparison was performed on histological data with mast cells. The dataset had regions with different types of borders. The borders of the regions varied from clearly visible to ragged, and in some cases, looked like a cloud of points.

For the semantic segmentation problem, ENet lost about 1-2\% of accuracy at the comparison with UNet in terms of the DSC and the F1-score. On the other hand, ENet computational performance is significantly higher (up to 10-15 times) that is important for many applications.

\bibliographystyle{IEEEtran}
\bibliography{bibliography}
\end{document}